\documentstyle[aps,twocolumn]{revtex}

\begin{document}


\title{Molecular Modeling of Trifluoromethanesulfonic Acid for
Solvation Theory}


\author{Stephen J. Paddison$^\ast$,
	Lawrence R. Pratt$^\dagger$,
	Thomas Zawodzinski$^\ast$, and David W. Reagor$^\ast$}
\address{Materials Science and Technology Division$^\ast$
	and Theoretical Division$^\dagger$
Los Alamos National Laboratory, Los
Alamos, New Mexico 87545 USA }

\author{LA-UR-97-1688}

\date{\today}

\maketitle

\def\rtrade{$\bigcirc\!\!\!\!\!${\tt R} }

\begin{abstract}Reported here are theoretical calculations on the
trifluoromethanesulfonic  (triflic) acid and water molecules,
establishing molecular scale information necessary to molecular
modeling of the structure, thermodynamics, and ionic transport of
Nafion\rtrade membranes.  The optimized geometry determined for the
isolated triflic acid molecule, obtained from {\sl ab initio} molecular
orbital calculations, agrees with previous studies.  In order to
characterize side chain flexibility and accessibility of the acid
proton, potential energy and free energy surfaces for rotation
of both carbon-sulfur and sulfur-oxygen(hydroxyl) bonds are
presented.  A continuum dielectric solvation model is used to
obtain free energies of electrostatic interaction with the solvent.
Electrostatic solvation is predicted to reduce the free energy
barrier to rotation of the F$_3$C-SO$_3$ bond to about 2.7
kcal/mol. This electrostatic effect is associated with slight
additional polarization of the CF bond in the eclipsed
conformation.  The energetic barrier to rotation of the acid
hydroxyl group away from the sulfonic acid oxygen plane, out into
the solvent is substantially flattened by electrostatic solvation
effects.  The maximum free energy change for those solvent accessible
proton conformations is less than one  kcal/mol.  We carried out
additional {\sl ab initio} electronic structure calculations with a 
probe water molecule interacting with the triflic acid.  The minimum
energy structures found here for the triflic acid molecule with the
probe water revise results reported previously.  To investigate the
reaction path for abstraction of a proton from triflic acid, we
found minimum energy structures, energies, and free energies for:
(a) a docked configuration of triflate anion and hydronium cation
and (b) a transition state for proton interchange between triflic
acid and a water molecule.  Those configurations are structurally
similar but energetically substantially different.  The activation
free energy for that proton interchange is predicted to be 4.7
kcal/mol above the reaction end-points.

\bigskip


\end{abstract}



\section{Introduction}

Nafion\rtrade, a perfluorinated ionomer widely used in polymer
electrolyte fuel cells \cite{Lemons}, consists of a hydrophobic
Teflon backbone with randomly attached side chains terminating with
trifluoromethanesulfonic (triflic) acid groups. These sulfonic acid
functional groups are hydrophilic and preferentially hydrated. The
Nafion\rtrade membrane functions as both a separator and an electrolyte
in the fuel cell and the overall performance of the fuel cell is
strongly influenced by the membrane conductivity, itself a function
of the hydration \cite{Zawod:91,Springer,Zawod:93a,Zawod:93b}. The
water content of the membrane is determined both by equilibrium
absorption and by nonequilibrium currents of ions and water. 
Extensive experimental measurement of electroosmotic drag coupling
proton and water fluxes for various sulfonated membranes over a
wide range of water content \cite{Zawod:93c,Zawod:95,Xie} has
suggested that sulfonic acid interactions dominate proton
transport.  A molecular level understanding of these processes in
Nafion\rtrade is not available.  Here we work towards a molecular
description of these systems by securing information on molecular
structures, energies, and charge distribution upon which molecular
scale solution theory might be based.

Electrostatic interactions of triflic acid with the environment are
expected to be of first importance in this solution chemistry.
Continuum dielectric solvation modeling is an appropriate approach
to studying these solution processes; it is physical,
computationally feasible, and can provide a basis for more
molecular theory 
\cite{Pratt:94,Tawa:94,Tawa:95,Hummer:95a,Hummer:96,Corcelli,%
Pratt:97,Hummer:97a,Hummer:97b}.

\section{Computational Methods}

\subsection{Electronic Structure Calculations}

Standard {\sl ab initio} molecular orbital calculations were
performed using the GAUSSIAN 92/DFT and 94 \cite{G92-DFT,G94}
systems of programs. Both the 6-31G** split valence basis set
\cite{Hariharan} and the D95** Dunning double zeta basis set
\cite{Dunning}  were employed at the Hartree-Fock plus second-order
M\o ller-Plesset (MP2) level of theory.  Full geometry optimizations
were obtained for isolated triflic acid (CF$_3$SO$_3$H), the
triflate anion (CF$_3$SO$_3^-$), and triflic acid with a probe
water molecule (CF$_3$SO$_3$H+H$_2$O) using gradient methods
\cite{Schlegel}. Atom centered partial charges were obtained by the
CHelpG scheme \cite{Breneman}.  Potential energy surfaces for
triflic acid were constructed from calculations at the MP2/6-31G**
level by rotation of the carbon-sulfur and
sulfur-oxygen(hydroxyl) bonds with optimization of all other
degrees of freedom.  Finally, the optimized structures and energies
were determined for the triflate-hydronium ion pair in two ways:
(a) by docking triflate and hydronium ions, each rigidly
constrained at their separate equilibrium geometries; and (b) by
finding a transition state for interchange of a proton in the
triflic-acid-water complex as CF$_3$SO$_3$H$^*$ + HOH =
CF$_3$SO$_3$H + HOH$^*$ using Synchronous Transit-guided
Quasi-Newton (STQN) methods \cite{Peng}.

\subsection{Dielectric Continuum Model}

A continuum dielectric model of aqueous solution chemistry has
become a standard theoretical tool in recent years and discussions
of basic issues can be found elsewhere \cite{Pratt:97}. 
The model identifies a
solute molecule in a solution environment, defines a realistic
solute molecular volume based on the geometry of the molecule,
positions partial charges describing the solute electric charge
distribution in that molecular volume, and assumes that the solvent
exterior to the molecular volume may be idealized as a dielectric
continuum. The numerical task is then the solution of the Poisson
equation for which a defined local dielectric constant jumps from
one (1.0) inside the molecular volume to the experimental value for
the solution in the region exterior to the solute molecular volume.
The numerical methods used here were boundary integral techniques
outlined in References \cite{Corcelli,Pratt:97}. The molecular volume 
is modeled
as the union of spherical volumes centered on solute atoms.  The
radii adopted for those spheres depend on atomic type and are
essentially empirical.  Here we used the values determined by
Stefanovich and Truong \cite{Stefanovich}.  The value 77.4 was used for the
solvent dielectric constant.

\section{Results and Discussion}

The optimized geometries of both triflic acid and the triflate
anion are essentially identical to the structures reported by
Gejji {\sl et al.} \cite{Gejji:93a,Gejji:93b} at the same levels of theory.  Both
structures are in good agreement with experimentally determined
structures \cite{Schultz,Delaplane}, the only difference being that the
sulfur-oxygen and carbon-fluorine bond lengths are 0.03-0.08\AA
longer in the theoretical structures. To characterize the
flexibility of the side chains of the perfluorinated membrane and
the accessibility of the acidic protons, potential energy surfaces
for triflic acid were determined for rotation of the
carbon-sulfur bond and the sulfur-oxygen(hydroxyl) bond.  The
relative energies for rotation of both of these bonds are
reported in Tables 1 and 2. The conformational potential energy
surface for rotation of the carbon-sulfur bond is given in
Figure 1. The highest energy structure occurs when the oxygen and
fluorine atoms are eclipsed and the minimum energy structure when
those atoms are staggered. The activation energy is 3.5 kcal/mol
but the solvation free energy lowers that barrier to about 2.7
kcal/mol.  This is mostly due to a slight additional polarization
of the C-F bonds in the eclipsed conformation. Figure 2 shows as
function of rotation of the carbon-sulfur bond the atom centered
partial charges obtained for the triflic acid molecule. The
sulfur-oxygen(hydroxyl) rotational potential energy surface
possesses two physically equivalent minima (Figure 3).  The lowest
barrier between these minima corresponds to the proton directed
axially, out into the solvent and has an energy of 2.0 kcal/mol.
Electrostatic solvation substantially flattens that barrier so that
the maximum free energy between the two minima is less than one
kcal/mol.  This stabilization of the barrier configuration is due
to the better access of the proton to the solvent. The position of
a probe water molecule interacting with triflic acid was determined
by fully optimizing the pair of molecules at the MP2 level using
both the 6-31G** and D95** basis sets. Recently, Ricchiardi and
Ugliengo \cite{Ricchiardi} reported a geometry optimization of triflic acid plus
a water molecule at the SCF/DZP level. Their result is
qualitatively similar but quantitatively substantially different
from that presented below. Because our results differed from those
previous ones, several different basis sets and starting conditions
were considered. The differences we observed among those
optimizations were slight and do not account for the differences
with the previous work that are likely traceable to
different level of effort in achieving the optimum structure. Our
results are shown in Figure 4. That structure is of lower energy
than the structure reported by Ricchiardi and Ugliengo.  However,
our experience with this optimization is that the
triflic-acid-water configurations are floppy as might be expected
from noncovalent hydrogen-bonding interactions.  For example, we
several times found a local minimum of the energy surface with the
exterior proton of Figure 4 directed into the plane of that
drawing.  Potential energy minima of that sort are only slightly
higher in energy, of the order of one kcal/mol, than the structure
shown.  Note that the dihedral angle positioning the triflic acid
proton is nearly the same in the minimum energy complex and the
minimum energy triflic acid molecule (Figure 3).  The floppiness of
these structures is consistent with the prediction of the
dielectric solvation model that free energy varies only gently with
disposition of the acid proton in the region between the two minima
(Figure 3). Attempts were made to determine a minimum energy
conformation for triflate-hydronium ion pair but those
configurations always relaxed to the optimum geometry of the
triflic-acid-water molecule pair.  Separately optimized structures
for the triflate anion and the hydronium cation were then docked in
an optimization that adjusted only the relative positions of the
two ions constrained separately to be rigid.  That optimum ion pair
geometry is presented in Figure 4 also.  The energy of the docked
ion pair is 27.0 kcal/mol higher than the fully optimized molecule
pair.  The electrostatic solvation free energy of the docked ion
pair relative to that of the optimized molecular complex is
computed to be -16.0 kcal/mol so that net free energy of the docked
ion pair, as a complex, is approximately 11.0 kcal/mol above that
of the minimum energy molecule complex. Further efforts to
determine a reaction path for abstraction of the proton from the
triflic acid molecule were based upon the observation (Figure 4)
that another minimum energy structure for the triflic-acid-water
complex is obtained by cycling the triflic acid proton with a
distinct water proton to achieve a physically equivalent but
different triflic-acid-water molecule pair.  This can be
alternatively described by considering that one of the two close
protons of the docked hydronium ion will be donated to the triflate
anion when it is neutralized.  We defined those two possibilities
as end-points for a reaction path and found a transition state
(Figure 4) using STQN methods [26]. Structurally, the docked ion
pair and the transition state are quite similar. The energy of this
transition state is 9.3 kcal/mol higher than the end-points of the
reaction path or about 18 kcal/mol lower in than the energy of the
docked structure.  The electrostatic solvation free energy relative
to the reaction end-point is predicted to be -4.6 kcal/mol so the
activation free energy is about 4.7 kcal/mol.

\section{Conclusions}

The results reported here establish molecular scale information
necessary to molecular modeling of the structure, thermodynamics,
and ionic transport of Nafion\rtrade membranes.  Because of a slight
additional polarization of the CF bond in the eclipsed
conformation, electrostatic solvation reduces the free energy
barrier to rotation of the F$_3$C-SO$_3$ bond to about 2.7 kcal/mol.
Solvation enhances the accessibility of the acid proton by
substantially flattening the energetic barrier to rotation of the
acid hydroxyl group away from the sulfonic acid oxygen plane, out
into the solvent.  The free energy barrier for those solvent
accessible proton conformations is less than one kcal/mol relative to
the minimum energy conformations. The minimum energy structures
found for the triflic acid molecule with the probe water 
revise structures reported previously.  A doubly
hydrogen-bonded configuration is found in which the triflic acid
proton forms a short hydrogen-bond. The second hydrogen-bond is of
a more traditional type. Minimum energy structures, energies, and
free energies were found for: (a) a docked configuration of
triflate anion and hydronium cation and (b) a transition state for
proton interchange between triflic acid and a water molecule. Those
configurations are structurally similar but energetically
substantially different.  The activation free energy for that
proton interchange is predicted to be 4.7 kcal/mol above the
reaction end-points. Experimentally determined values of activation
energies for protonic conductivity in Nafion\rtrade and similar
perfluorosulfonic acid membranes are roughly 3 kcal/mole for a
fully hydrated membrane and increases with decreasing water content
[5].

\bigskip
\bigskip
\bigskip
\vfill
\appendix{{\bf Acknowledgements}}

We thank P. J. Hay and R. L. Martin for helpful discussions.  This
work was supported by the Los Alamos National Laboratory LDRD
program.

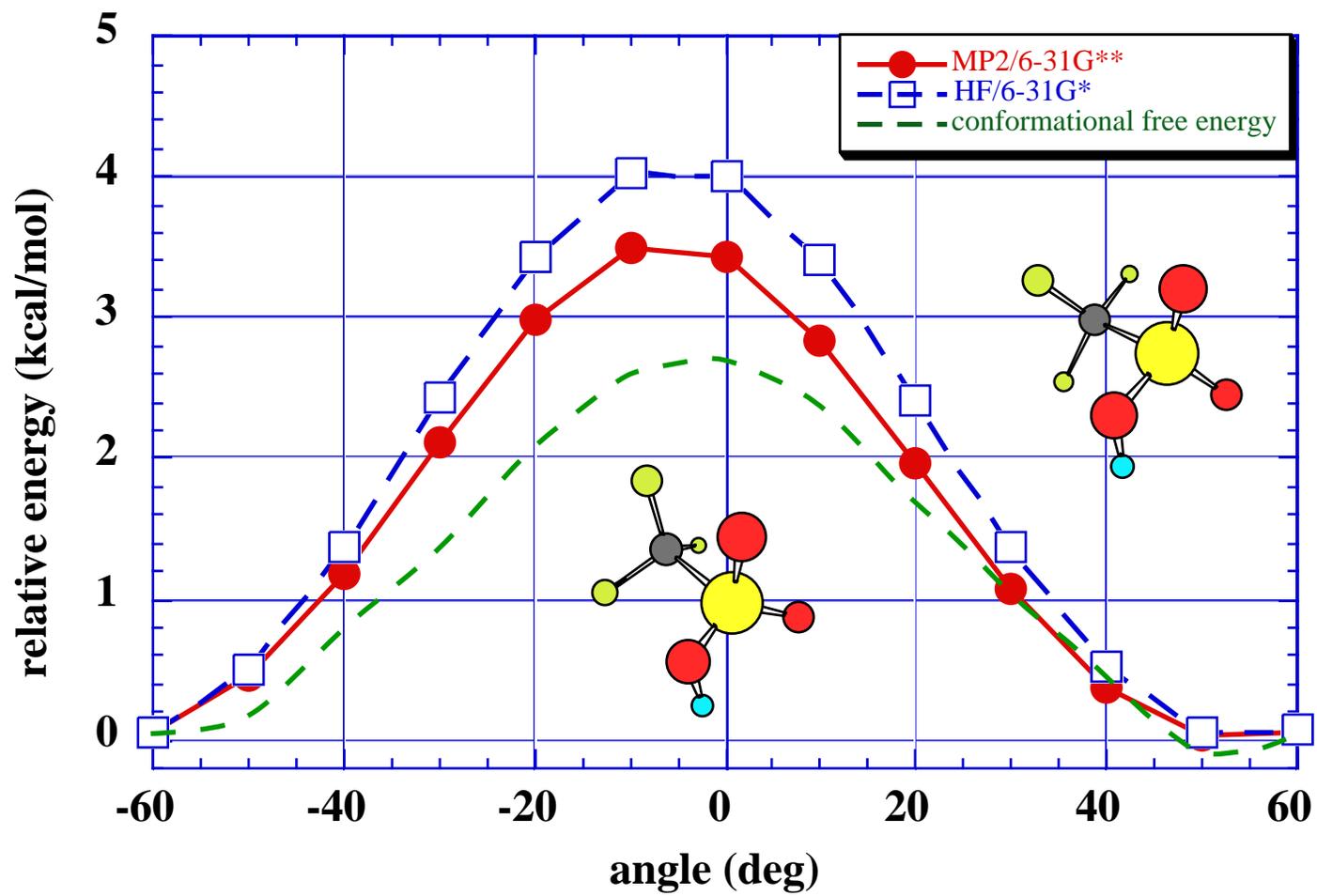
\begin{figure}

\caption{Relative potential energy as function of rotation of  
the carbon-sulfur bond 	in triflic acid.  The insets
show the molecular geometries for the eclipsed (left, barrier
maximum) and staggered (right, minimum) conformations.}

\label{fig1}
\end{figure}

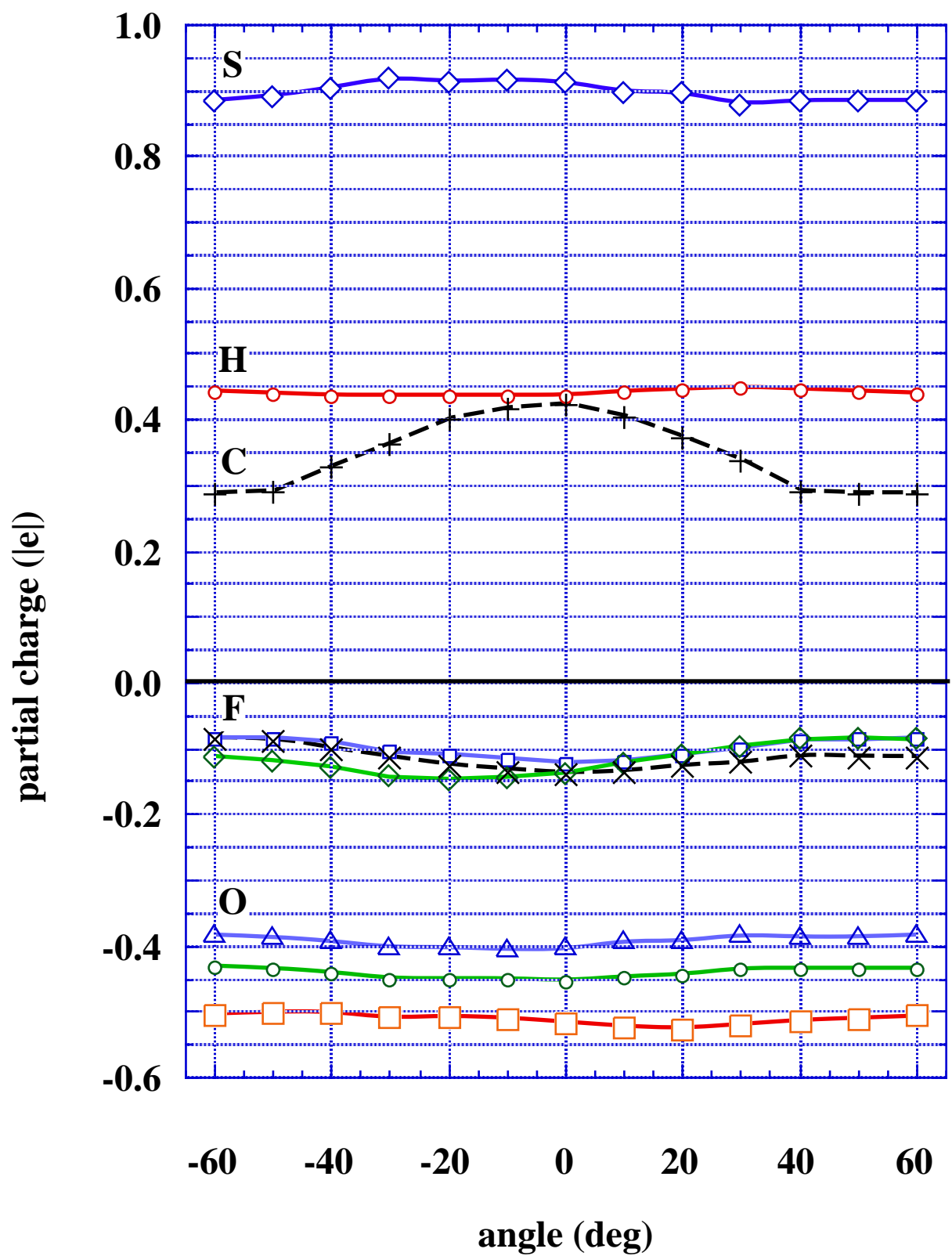
\begin{figure}

\caption{Atom centered partial charges for the triflic acid
molecule as a function of the rotation of the
F$_3$C-SO$_3$ bond.}

\label{fig2}
\end{figure}

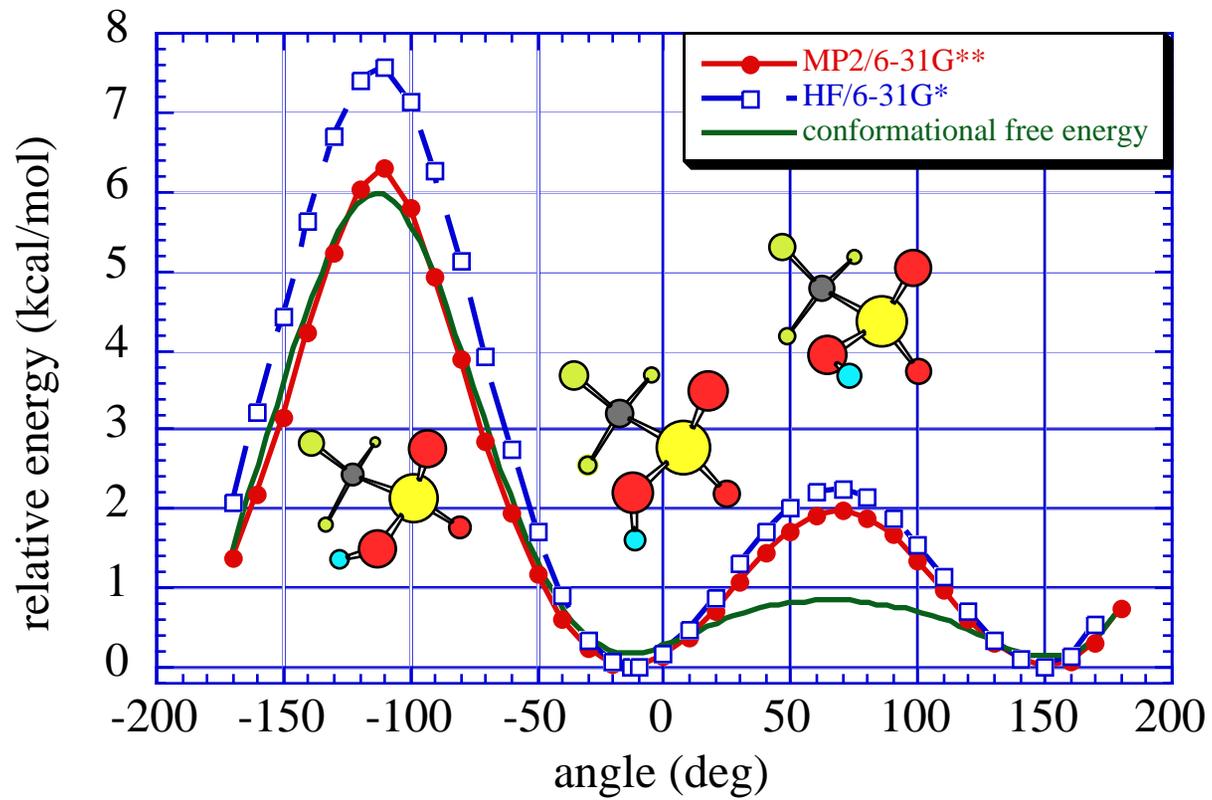
\begin{figure}

\caption{Relative potential energy as function of rotation of  
the sulfur-oxygen (hydroxyl) bond in triflic acid. The insets
show the molecular geometries for the successive extrema
starting at the left from the maximum energy.}

\label{fig3}
\end{figure}

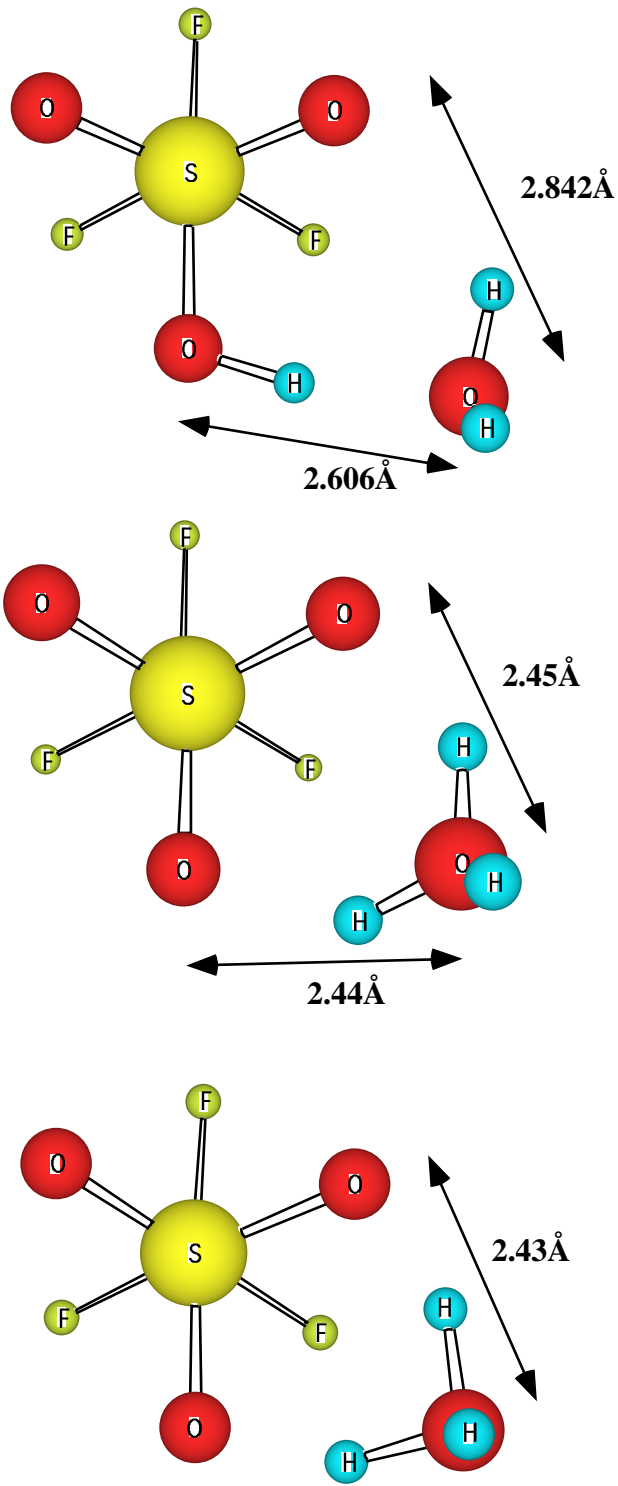
\begin{figure}

\caption{(upper) MP2/6-31G** optimized geometry of the
triflic-acid-water molecule pair. (middle) Docked
structure geometry of the triflate anion and hydronium
cation pair. The free energy of this structure is
about 11 kcal/mol higher than that of optimum
molecular complex above. (lower) STQN Transition structure
CF$_3$SO$_3$$^-$+H$_3$O$^+$.  The free energy of this structure is
about 4.7 kcal/mol higher than that of upper
configuration.}

\label{fig4}
\end{figure}

%
%

\begin{table} 
\caption{Triflic acid relative potential energy for rotation
of the F$_3$C-SO$_3$ bond} 
\label{Table 1} 
\begin{tabular}{dd}
angle & MP2 relative energy \\
(deg) & (kcal/mol) \\ \hline
60.0 & 0.07 \\
50.0 & 0.03 \\
40.0 & 0.38 \\
30.0 &	1.07 \\
20.0 &	1.96 \\
10.0 &  2.84 \\
0.0 &  3.44 \\
-10.0 & 3.50 \\
-20.0 & 2.99 \\
-30.0 & 2.12 \\
-40.0 & 1.19 \\
-50.0 & 0.45 \\
-60.0 & 0.05 \\
-65.42 & 0.00 \\
\end{tabular} 
\end{table}

\begin{table} 
\caption{Triflic acid relative potential energy
for rotation of the sulfur-oxygen (hydroxyl) 
bond
} 
\label{Table 2} 
\begin{tabular}{dd}
angle & MP2 rel ative energy \\
(deg) & (kcal/mol) \\ \hline
180.0 & 0.74 \\
170.0 & 	0.31 \\
160.0 & 	0.07 \\
150.0 & 	0.00 \\
140.0 & 	0.09 \\
130.0 & 	0.31 \\
120.0 & 	0.62 \\
110.0 & 	0.98 \\
100.0 & 	1.34 \\
90.0 & 	1.66 \\
80.0 & 	1.89 \\
70.0 & 	1.98 \\
60.0 & 	1.92 \\
50.0 & 	1.73 \\
40.0 & 	1.43 \\
30.0 & 	1.07 \\
20.0 & 	0.71 \\
10.0 & 	0.38 \\
0.0 & 	0.14 \\
-10.0 & 	0.01 \\
-20.0 & 	0.03 \\
-30.0 & 	0.23 \\
-40.0 & 	0.61 \\
-50.0 & 	1.18 \\
-60.0 & 	1.94 \\
-70.0 & 	2.87 \\
-80.0 & 	3.90 \\ 
-90.0 & 	4.93 \\
-100.0 & 	5.79 \\
-110.0 & 	6.28 \\
-120.0 & 	6.02 \\
-130.0 & 	5.23 \\
-140.0 & 	4.23 \\
-150.0 & 	3.17 \\
-160.0 & 	2.19 \\
-170.0 & 	1.37 \\
-13.52 & 	0.00 \\
\end{tabular} 
\end{table}
\end{document}